# Superconducting Qubits as Musical Synthesizers for Live Performance

Spencer Topel[1,2], Kyle Serniak[1,3]*, Luke Burkhart[1,4], and Florian Carle[1]*

[1] *Yale Quantum Institute, Yale University, New Haven, CT, USA*
[2] *Physical Synthesis, Brooklyn, NY, USA*
[3] *Current Address: MIT Lincoln Laboratory, Lexington, MA, USA*
[4] *Current Address: Keysight Technologies, Cambridge, MA, USA*
*Corresponding authors: florian.carle@yale.edu, kyle.serniak@ll.mit.edu

**Abstract**

In the frame of a year-long artistic residency at the Yale Quantum Institute in 2019, artist and technologist Spencer Topel and quantum physicists Kyle Serniak and Luke Burkhart collaborated to create *Quantum Sound*, the first-ever music created and performed directly from measurements of superconducting quantum devices. Using analog- and digital-signal-processing sonification techniques, the team transformed GHz-frequency signals from experiments inside dilution refrigerators into audible sounds. The project was performed live at the International Festival of Arts and Ideas in New Haven, Connecticut on June 14, 2019 as a structured improvisation using the synthesis methods described in this chapter. At the interface between research and art, *Quantum Sound* represents an earnest attempt to produce a sonic reflection of the quantum realm.

*The text is an unedited pre-publication version of a chapter which will appear in the book "Quantum Computer Music," Miranda, E. R. (Editor), MIT Press.*

# 1   Introduction

The past several decades have seen breakthroughs in both the theory and the practice of quantum science. The quantum phenomena of superposition and entanglement are now understood as unique resources that can be harnessed to solve computationally intensive problems. Experimental progress has enabled precise control over individual quantum objects, whether naturally occurring microscopic systems like atoms, or carefully designed macroscopic quantum systems whose properties can be engineered.

These advances may soon allow us to perform otherwise intractable computations, ensure privacy in communications, and better understand and design novel states of matter. The emerging discipline of quantum information processing combines physics, electrical engineering, mathematics, and computer science to further the basic understanding of the quantum world, and to develop novel computational devices and other quantum-enabled measurement and sensing technologies.

The speed of development of quantum technologies has increased dramatically over the last few years. The prospect of novel applications of quantum computers has triggered a frenzy of investment and intense development in a race toward the first demonstration of quantum advantage, wherein a quantum computer definitively outperforms a classical computer. But in this race, most people outside of the field of quantum computing are left wondering how this technology works and how it might impact their lives. With the goal of encouraging public conversation about quantum science and to combat scientific reticence, Florian Carle created an Artist-in-Residence program for the Yale Quantum Institute in 2017. This program brings talented artists into the research laboratories to produce artwork in collaboration with quantum physicists. By exploring art as a medium and leveraging the intersectionality of science and the humanities, this program increases public understanding and discourse of quantum physics by seeing this topic from the perspective of a collaborative artist. In building this program, an emphasis was put on connecting the right people with the proper expertise at the right time to make something truly unique and inspiring to a new generation, and to give them the desire to push the boundary of our collective knowledge.

In this chapter, we will focus on the live performance of *Quantum Sound* – a collaborative work of the second Artist-in-Residence, Spencer Topel, who is a co-author on this publication. Spencer is a musician, artist, and researcher working with electromechanical music synthesizers, interactive sound installations, live performances, and recordings. In 2019, Spencer collaborated with Kyle Serniak and Luke Burkhart, two quantum physicists finishing their PhDs at Yale University to "play" sounds generated by the operation of prototype quantum processors in the Yale Quantum Institute laboratories, cooled to nearly absolute zero (~10 millikelvins), as if they were instruments in an orchestra. This live performance was the first of its kind, attracting the interest of scientists in the quantum community due to its technical challenge, and of the public for the novel soundscape that was produced.

In many respects, quantum physicists and electronic musicians speak a similar language, which is why there has been a surge of interest in quantum computer music in the last decade with projects like *Quantum Music* [1,2] or *Quantum Sound* [3], the creation of quantum-music-centric conferences [4,5], or even books like this one. Musicians and researchers interact with their instruments thinking in common terms such as frequencies, signals, and standing waves. Furthermore, they must worry about similar problems like electromagnetic interference, noise, and aberrant resonances. Keeping in mind that the advent of personal computing served to democratize electronic music production, it's worthwhile to assess whether or not quantum technologies will further expand the musician's toolkit. Though we do not address this question directly, our demonstration – working with fundamental technologies being developed in pursuit of larger-scale quantum systems – echoes experimentations with new musical processes from more than a century ago, when Leon Theremin invented instruments and transducers exploiting principles of electromagnetism and recently-invented radio proximity sensors [6]. The most famous of these instruments, the theremin, became the first contactless musical instrument and is still fabricated and performed today.

This chapter discusses a mixture of science and art, as it relates to sonification (the act of expressing inaudible phenomena as sound) and synthesis intended to describe to the reader (in language meant to bridge the two fields) the tools we utilized to create music using measurements taken from quantum systems. Section 2 will give an overview of the technology and hardware used to generate the data. Section 3 will focus on the synthesizer programming and design, and how the signals were made usable and audible. Section 4 will describe the live performance, detail the musical motifs therein, and explain the artists' intentions for the piece.

## 2  Quantum Experiments Used to Generate Sounds

Quantum information processing relies on precise control of quantum systems. In recent years, groundbreaking quantum information experiments have solidified the position of superconducting quantum processors as a viable technology for the Noisy Intermediate-Scale Quantum era and beyond [7,8] (NISQ, referring to the stage where quantum processors are reaching 100's of qubits and can be utilized for small-scale, specifically tailored algorithms). While atoms and subatomic particles are archetypes of the quantum world, some solid-state systems can also behave quantum mechanically. In the case of superconducting qubits [9,10] the "quantumness" is observed in the dynamics of collective excitations in electrical circuits [11] when they are cooled far below the superconducting transition temperature (for aluminum, the most common material used in the field, this transition occurs at ~1.2 kelvins, or ~-457 degrees Fahrenheit). In these systems, relevant transition frequencies are in the few GHz range, which translates to energies greater than the temperature at which experiments are conducted. This is most commonly facilitated by cryogenic systems (commonly referred to as "cryostats") called dilution refrigerators (see *Figure 1*), which nowadays can be purchased commercially and have become a ubiquitous tool for low-temperature physics. [12–14]. These systems can operate at ~10 millikelvins, which is nominally cold enough to freeze out all electronic excitations in the superconductor and have enough cooling power to overcome the power dissipation associated with control and measurement of many superconducting qubits.

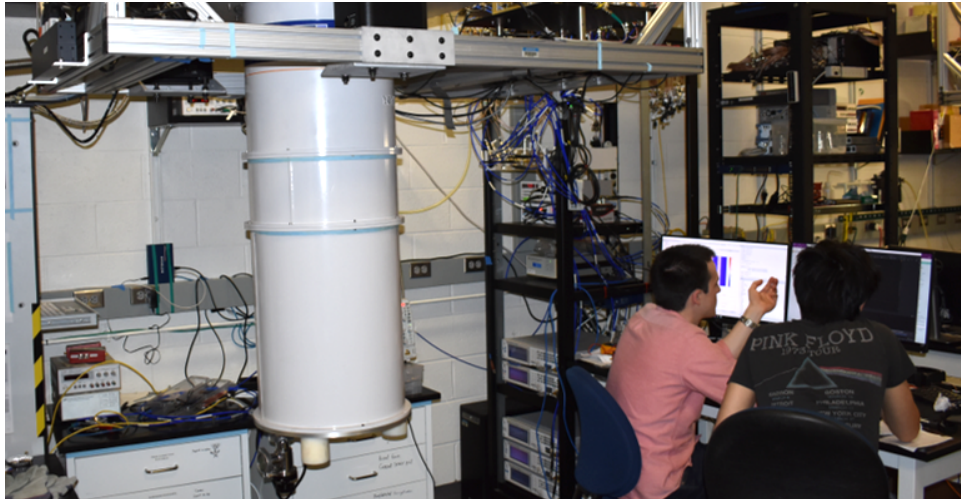

*Figure 1 -* *Photograph of Blue, one of the two dilution refrigerators that housed the experiments used as physical synthesizers to perform Quantum Sound, and Luke Burkhart (left), one of the performers.*

The qubit devices addressed during the performance of *Quantum Sound* were designed for specific experiments related to superconducting quantum information processing. Each are based on "transmon" qubits [8], which consist of two circuit elements in parallel: a capacitor and a uniquely superconducting component – the Josephson junction (JJ). In the transmon, the JJ serves as a nonlinear inductor and is crucial to achieving coherent control of the qubit. Transmon qubits are constructed using traditional lithography and deposition techniques, similar to those used in semiconductor circuit manufacturing (see *Figure 2*).

Experiments performed using superconducting qubits rely on low-noise electrical signal generation ranging from DC to around 10 GHz, surpassing the audio-frequency range by many orders of magnitude. Passive and active components up to these frequencies are readily available commercially due to vast research and development efforts related to telecommunications and radar. These established technologies facilitate high-fidelity control of qubit states with microwave-frequency drives, which means that experimentalists can prepare the qubit in any superposition of its energy eigenstates with error rates approaching 1 in 10000 [15].

With the exception of direct qubit control explicitly utilized in Luke's experiment for *Quantum Sound*, the qubit-state readout was responsible for most of the musical choices made in the performance. Whether a qubit is occupying its ground or excited state is determined via a technique called dispersive readout, in which the qubit state is detected via a state-dependent frequency shift of a nearby harmonic oscillator, which is also made from superconducting materials (henceforth referred to as a transmon-cavity system). The qubit's state is encoded in the amplitude and phase (or in Cartesian space, the in-phase and quadrature components) of a microwave signal that interacts with the oscillator. Therefore, by monitoring that signal, one can efficiently determine the state of the qubit. With quantum-noise-limited parametric amplifiers, sufficient signal-to-noise (SNR > 1) for single-shot readout can be achieved with measurement durations of a few 100's of nanoseconds, the inverse of which sets the maximum sampling rate with which signals in our experiments could be sonified. These signals are mixed down from GHz to ~10s of MHz using standard frequency mixing techniques, then detected with a high-bandwidth analog-to-digital converter and demodulated.

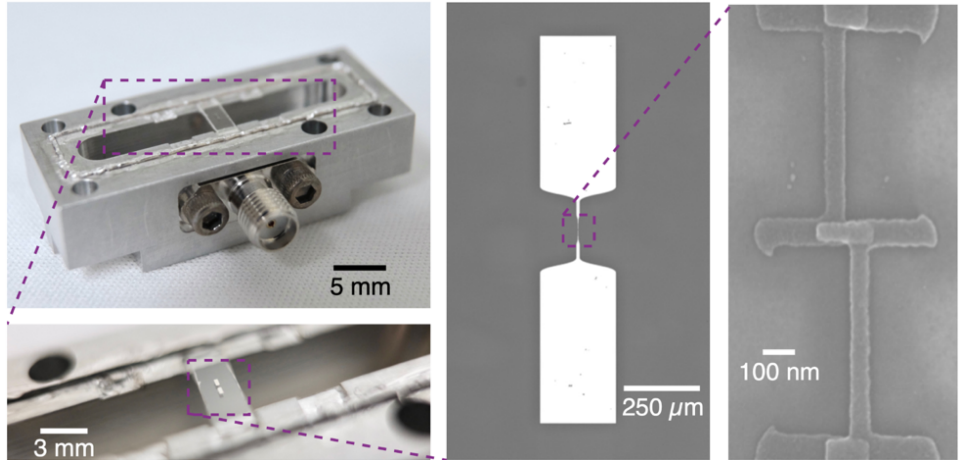

*Figure 2 - Physical realization of a 3D transmon superconducting qubit. Top left shows a photograph of one half of the 3D waveguide cavity resonator that houses the transmon and is used to readout the state of the qubit (in this case, it was machined from 6061 Al, a superconducting alloy). Zooming in, the bottom left shows the sapphire substrate, upon which the transmon was fabricated, mounted in the center of the cavity. Zooming in further, the center panel shows an optical micrograph of the transmon, with large coplanar capacitor paddles. On the right is a scanning electron micrograph (SEM) of an Al/AlOx/Al Josephson tunnel junction fabricated using the "bridge-free" technique [16]. Figure reprinted from Kyle Serniak's PhD Thesis [17].*

While each experiment discussed here [17,18] used similar hardware, the motivations for these experiments differed significantly, illustrating the breadth of current research on superconducting qubits. The first experiment performed by Kyle Serniak focused on readout and noise characterization. The second experiment performed by Luke Burkhart focused on high-fidelity qubit control based on classical feedback. The following sections summarize these experiments, whose data provided material for the *Quantum Sound* performances.

## 2.1 Nonequilibrium Quasiparticles in Superconducting Qubits

Kyle Serniak's experiment focused on understanding a particular mechanism of errors in superconducting qubits – namely those from nonequilibrium quasiparticle (QP) excitations [19] QPs are electronic excitations in a superconductor and are why it is so important to operate at very low temperature. So-called "BCS" superconductors [20], named after the physicists Bardeen, Cooper, and Schrieffer for their development of this theory, are characterized by a phonon-mediated electron-electron interaction that produces an energy gap in the electronic conduction band centered at the chemical potential. This gap serves as the backbone for a plethora of mesoscopic quantum phenomena (in the case of aluminum, this gap around 400 micro-electronvolts) in that it suppresses electronic excitations (the aforementioned QPs) at low energies. At low temperatures there is an exponentially small occurrence of thermally generated excitations above this gap. In fact, in order to find just one pair of excitations in an otherwise isolated block of aluminum sitting at 10 millikelvins one would *expect* to need a volume of material larger than the Earth! Unfortunately, coupling to external sources of energy such as infrared radiation and radioactive decay products results in an unexpectedly high, *nonequilibrium* population of QP excitations.

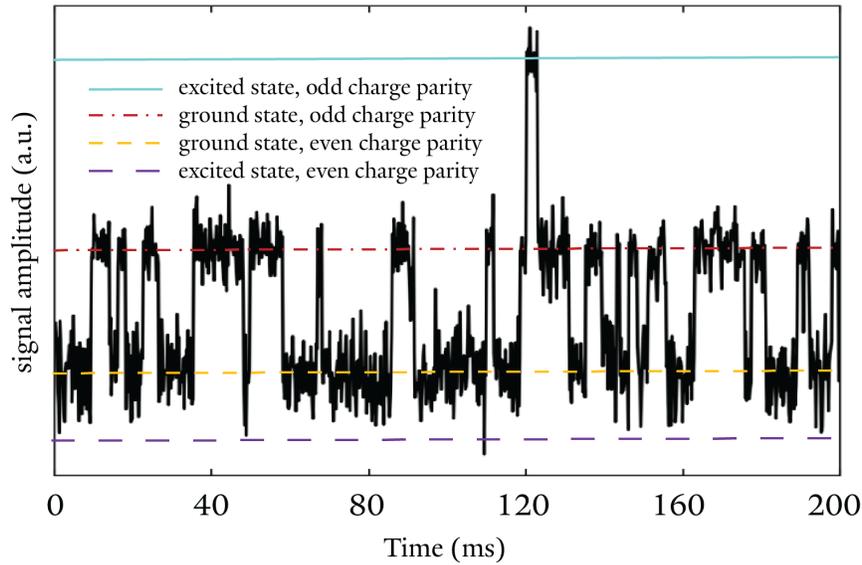

*Figure 3 - Simultaneous measurement of qubit state (ground or excited) and charge parity (the number parity of quasiparticles (QPs) that have tunneled across the Josephson junction, even or odd) in a transmon qubit [19]. Experimental data is shown with overlaid horizontal lines denoting the signal amplitudes that correspond to each state. Signals such as these were the basis for sounds generated for the performance and sonified using the 4-state synthesizer (see Section 3.2).*

These QPs will tunnel across the JJ of the transmon, and when they do, they can induce qubit errors. To characterize the rate of these errors and distinguish them from other error mechanisms, it's possible to monitor other observables in the transmon called the *offset charge* and *charge parity*. The former is a classical voltage difference across the JJ due to spurious charges in the environment that reconfigure on a timescale of a few minutes, and the latter of which is the number parity (even or odd) of QPs that have tunneled across the JJ (which tunnel on a timescale of a few milliseconds). The goal of this experiment was to characterize the timescale on which QP tunneling events occurred in transmon qubits, and to do so in a way that is more efficient than previous demonstrations [21,22]. Using a single transmon-cavity system, this experiment utilized the same dispersive coupling responsible for qubit-state readout to also measure the charge parity of the transmon. The result was a readout signal from which one could determine not only the state of the qubit (ground or excited), but also the charge parity (even or odd) simultaneously (see *Figure 3*). Repeated measurements of the system, acquired with a 5 kHz-100 kHz sample rate, were used as waveforms to be processed into sound in the *Quantum Sound* performance.

## 2.2 Error-Detected Networking for 3D Circuit Quantum Electrodynamics

The other experiment, performed by Luke Burkhart, aimed at generating quantum entanglement between physically separated but linked quantum systems (see *Figure 4*) [23]. Although no aspect of this entanglement was explicitly used in the performance, key technological primitives necessary to operate these quantum systems were put to the test in order to create various sonic motifs. The device consisted of two transmon-cavity systems coupled via an auxiliary cavity mode. Each transmon-cavity system was capable of independent readout and control, which was used to musical effect, creating waveforms that were "shaped" by deterministically controlling the state of the qubits. Two examples of this included 1) applying $X(\pi)$-gates to a single qubit conditioned on fast feedback of its own state (practically referred to as state initialization or reset), and 2) applying $X(\pi)$-gates to one qubit conditioned on fast feedback of the *other qubit's* state. Errors in this feedback protocol (the so-called "Bad Follower" synthesis engine) were introduced by intentionally over- or under-rotating the "follower" qubit, effectively enacting $X(\theta)$-gates for continuous $\theta$ in $[0, 2\pi)$. All measurements were performed in the $\sigma_z$ basis, and therefore these suboptimal gates effectively simulated imperfect state preparation in the computational basis.

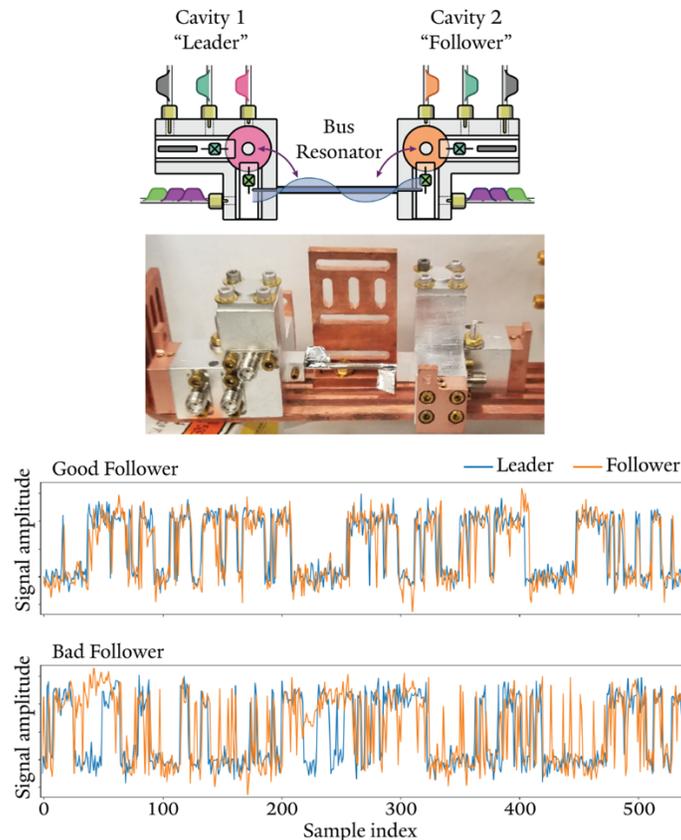

***Figure 4*** - *Diagram [18] and photograph of Luke's two qubit-cavity device (top). Cavity 1 ("leader") is coupled to Cavity 2 ("follower") via a cavity bus. The qubit housed in Cavity 2 is able to track the state of that in Cavity 1 via high-fidelity control and classical feedback ("Good Follower," top plot). By intentionally introducing errors in the control sequence, the correlation between readout signals can be reduced ("Bad Follower," bottom plot).*

Similarly, modulation of the expected qubit-state population as a function of time was achieved by modulating the rotation angle, creating waveforms similar to those of voltage-controlled oscillators in analog synthesizers, with imperfections induced by the stochastic, digital nature of projective measurements (see *Figure 5*).

## 2.3   Experimental Data as Control Voltages

Measurements of transmon qubits are represented by voltage outputs from analog-to-digital converters (ADCs). Repeated measurements (potentially with high sample rates up to ~1 MHz) strung together sequentially produce signals that can be translated to audio-rate signals relatively simply via resampling. The jumps in these signals represent state changes (see *Figure 5*), which at the simplest level can be used as markers to "select" or trigger notes, tones, or other musical events. Alternatively, the amplitude of the signal can be used directly as an equivalent to control voltage (CV) generation in traditional analog synthesizers.

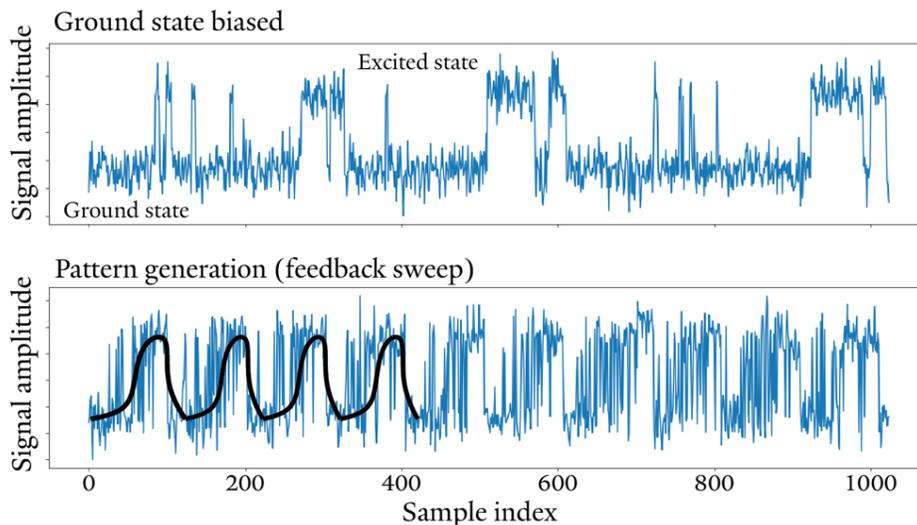

*Figure 5* - *Some synthesis tools/functions created to generate diverse sounds. Top: qubit dynamics with fixed feedback controls, biased toward ground-state occupation. Bottom: sweeping feedback parameters to modulate the expected qubit-state occupation.*

These signals also contain a certain amount of noise from of quantum fluctuations, which is detectable due to the use of nearly-quantum-limited parametric amplifiers [24,25]. While this could be used as a stochastic musical source, they can also be used to modulate oscillators creating pitch vibrato and other kinds of audible musical effects. Using quantum fluctuations as a modulating signal is one example of direct sonification of quantum phenomena.

## 3   A Simple Quantum Synthesizer

In order to sonify the signals acquired from the aforementioned experiments, we developed a simple quantum synthesizer utilizing measurement traces as an input. A block diagram that illustrates this synthesizer is shown in *Figure 6*. The majority of synthesizer designs used in the performance are variations of this model.

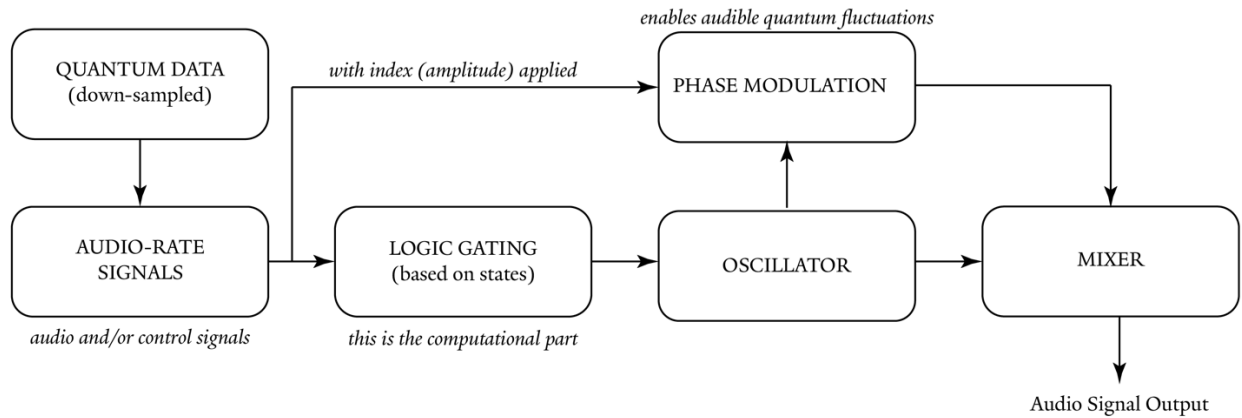

*Figure 6 - Process block diagram for a simple quantum music synthesizer. The experimental data is first resampled to audio rate, and then utilized to control oscillator parameters selected by the user.*

The basic concept is to take the output from the quantum systems and down-sample them to audio rate with a DC offset to center the signal at 0V, scaled to 2V pole-to-pole. These processed signals act as an oscillator operating in audible range (kHz) with the qubit-transition information used as a primary control voltage and the quantum noise signal extracted for modulation of the oscillator as a secondary effect. A modulation index parameter allows the user to control the amount of modulation applied, which uses the quantum noise encoded in the signal to generate its waveform. An index equal to zero would result in no modulation effect applied, with increasing index corresponding to an increasing modulation effect.

For the performance and album of *Quantum Sound,* we designed three synthesizers based on the simple model of the quantum synthesizer described above. In this section we will discuss each model and the resulting synthesizer design, with considerations for what was included as controls and what could be added in future iterations. The three synthesizers we designed were based on the experiments summarized in *Section 2*, and are identified as follows:

1. *2-State* is a two-state synthesizer with selectable waveforms, amplitudes, and frequencies for each state that can be phase modulated with the quantum fluctuations captured in the measurements taken from the aforementioned experiments.
2. *4-State* is a four-state synthesizer with similar features to *2-State:* selectable waveforms, amplitudes, and frequencies for each state that can be phase modulated with the quantum fluctuations captured in the measurements taken from the aforementioned experiments.
3. **Bad Follower** is a leader-follower synthesizer whereby the state-following behavior is accurate or inaccurate depending on the fidelity of classical feedback controlling the state of the "follower" qubit based on measurement outcomes of the "leader" qubit.

In addition to these synthesizers, we also created a synth engine built into both *2-State* and *4-State* that sonifies a measurement of the offset charge in Kyle's transmon qubit as a function of time, which due to its slow time dynamics could be played back at near its native sampling rate. Colloquially we referred to this as "drift." The drift synth engine used the normalized offset charge as a continuously-valued control voltage that determined the pitch of a musical drone produced by a sine-wave oscillator. To reduce the number of interfaces we were performing with, we added the drift sonification to *2-State* and *4-State*, which is read from a separate datafile, though it is worth noting that this sonification process could be a stand-alone synthesizer by itself.

### 3.1  *2-State*

The *2-State* synthesizer is a realization of the synthesizer model described above. We created this synthesizer in the Max-for-Live Ableton environment [26] since it is a flexible and fast way to create customized synthesis processes with a high degree of usability. The input to the system is a monophonic audio file containing the experimental data down-sampled to audio rate. A latching-filter state assignment is performed, which determines which oscillator is selected. The user can select a number of standard synthesis parameters for each state/oscillator: waveform type; frequency (in Hz); attack and decay envelope of; and modulation amount. The user can also control global parameters such as overall gain output from the synthesizer, data smoothing to reduce noise, and the wet/dry mix which results in either the raw data sonification, the processed data as synthesized tones, or some mixture of the two.

### 3.2  *4-State*

*4-State* expands the number of states from two to four enabling the user to create four-note melodic patterns based on the experimental data described in *Section 2.1*. Like *2-State,* the synthesizer is activated through a monophonic audio file containing the experimental data. Both interfaces for 2-State and 4-State contain similar user controls shown in *Figure 7*.

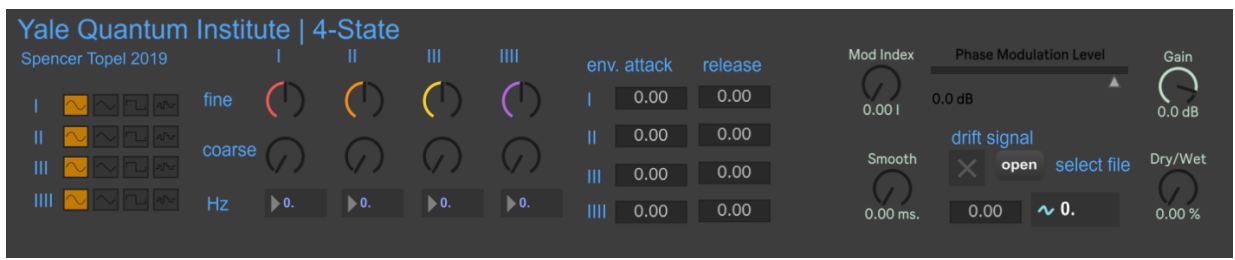

*Figure 7* – *Ableton Max-for-Live virtual device interface for 4-State. The system allows the user to set different waveforms for the four oscillators (sine, saw, square, and noise), the pitches of these oscillators and their basic envelopes (attack and decay). The controls for modulation driven by quantum noise include modulation index and modulation amplitude. A secondary process was performable from this interface, which allowed for the sonification of offset-charge drift (shown bottom right).*

## 3.3 *Bad Follower*

*Bad Follower* explores experimental methods described in Section 2.2, whereby the investigators utilize a two transmon-cavity device with independent readout and control producing a leader-follower arrangement, represented as two separate audio signals, left and right, respectively. Depending on the parameterization of the system, the state of the follower qubit would either accurately follow or inaccurately follow the state of the leader qubit using fast, classical feedback. The resulting data was converted to audio files, allowing the user to select good, average, or bad following, which is clearly audible when smoothing is applied. Like the other synthesizers described in this section, the user can apply phase modulation to both the data channels. A smoothing parameter can also be applied to both channels to produce a more continuous melodic pattern.

# 4 *Quantum Sound*: Superconducting Qubits as Musical Synthesizers

*Quantum Sound* premiered on June 14, 2019, at Firehouse 12, a jazz concert hall and recording studio during the 24[th] International Festival of Arts and Ideas of New Haven, CT (USA). Spencer, Kyle, and Luke performed two back-to-back 35-minute musical sets, preceded by a 15-minute introduction to the quantum physics behind the performance, and followed by a 10-minute question-and-answer session between the audience and the artists.

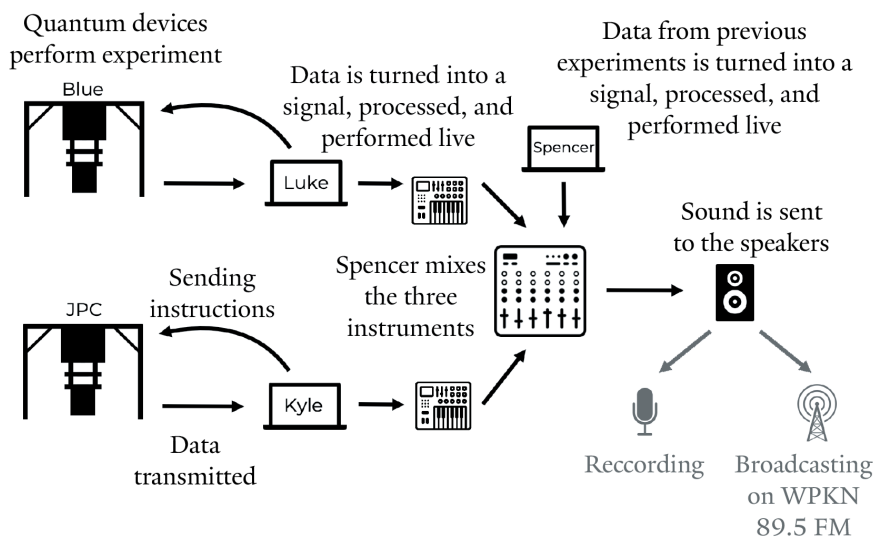

*Figure 8* - *Diagram of data acquisition and process during the live performance to transform quantum signals into an audible musical experience.*

During the performance, Kyle and Luke used data acquired (some pre-recorded, some acquired live during the show) from two experimental setups (the dilution refrigerators affectionately named "Blue" and "JPC," located back in the laboratories on the 4[th] floor of Becton Center), which was processed and performed live (see *Figure 8*). Spencer, using pre-recorded data, processed and performed that data while mixing the three audio signals like a conductor to create a coherent composition. The second set was broadcast live on WPKN Radio to allow as many people as possible to experience the performance.

It is difficult to illustrate in writing[1] the variety and range of sounds produced by the superconducting instruments. The performance includes moments of intense noise, intercut by melodies that are at different times diatonic, ominous, or almost imperceptible. The audience, safely protected inside the bent-wood cocoon of Firehouse 12, was transported to an eerie and desolate winter landscape, evoking sounds akin to winds blowing through a mountain pass as well as looming storms. At times there is a comforting flutter bringing warmth to the listener, which Spencer calls a "wobble," created by the quantum noise that exists within the sonified data. The completed *Quantum Sound* composition resides at the intersection of polyrhythmic electronica and ambient noise music.

Lighting, designed by Florian, played an important role in the performance. Composed of solid and intense colors, the light evolved as the performance (and system) transitioned from a calm section (blue), to a fast paced one (red, *Figure* ), with a transition period with reds and blues mixing in a superposition of states with shifting pinks, mimicking line-cuts of Wigner quasiprobability distribution diagrams [27]. The intentional lack of green or any natural lighting echoed the nearly complete isolation of superconducting qubits deep inside dilution refrigerators at extremely low temperature.

## 4.1 From Noise to Meaning

In order to stay true to the notion of scientific outreach via musical performance, it was of utmost importance that the artistic choices taken throughout be relatable back to aspects of the experiments themselves. Unfortunately, the full relation to experiment was not something that was particularly accessible *during* the performance, but the audience picked up on many motifs that were discussed during the subsequent question-and-answer sessions. Here we seek to explain for the record some of these relations, and how certain motifs were inspired by aspects of the experiments.

As a whole, the structure of the performance can be considered as four distinct sections:

1. Noise (primitive, raw, direct)
2. Rhythmic (randomness into patterns)
3. Crescendo (melodic into danger)
4. Release (ambient resolution)

The first section of the piece is characterized by faithful representation of the collected data as waveforms which are played at the natural sampling rate of the experiment (between 1 kHz and 100 kHz), resulting in an atmosphere of unrecognizable noise. After the three "instruments" are introduced by entry of the performers, artistic liberty was taken to shape the noise into something akin to a conversation between the instruments. Various filtered noise voices enter and exit with movement arising from manual sweeps of filter bandwidth and cutoff frequency. This is not dissimilar to how noise is intentionally filtered via trial and error of different hardware components in experimental setups to achieve optimum performance of the quantum systems. *In situ* filtering of the noise that decoheres a quantum superposition state (known as dynamical decoupling [28]) is another example of noise shaping in experiments.

---

[1] NOTE: Readers can listen to the album *Quantum Sound* at QuantumInstitute.yale.edu/Quantum-Sound as well as watch a mini live performance inside the Yale quantum laboratories next to the quantum computer which generated the data.

Discernable pitch is introduced gradually as a transition into the second section of the performance. The simplest approach to creating pitched melodic structure was to feed a synthesis engine with a normalized measurement record where signal amplitude is translated to pitch. With this approach, a given pitch corresponds to a particular quantum state, and vibrato (subtle variations of pitch) was a direct representation of amplified quantum noise. A more "digital" synthesis method was also utilized, in which a latching-filter state assignment was performed on a qubit measurement trace, discretizing the audible pitches. Examples of these techniques can be heard as many "instruments" in the second section. A staccato line evocative of Morse code was created from a short snippet of resampled data in which the pitch corresponding to the qubit ground state was inaudibly low, while the excited state was set to around 700 Hz pitch.

Occasional "pings," akin to the way active sonar from submarines is presented in movies, marks the entrance of sustained tones. For these, a resampled and state-assigned measurement trace served as a volume gate in addition to determining pitch. The pings were triggered by occupation of the excited state of the transmon, with two audible pitches denoting the transmon's two charge-parity states. These qubit excitations are relatively rare, as the excited state was only populated ~6% of the time in this particular experiment.

The second section makes heavy use of this state assignment (both from the charge-parity mapping experiment and the single-qubit control experiment) as a CV to determine the pitch played by each instrument as the pace quickens and a "bassline" emerges. While the first section of *Quantum Sound* is arguably the truest sonic representation of the data, this second section is truest to the *interpretation* of the data, as experimentalists oftentimes rely on state assignment of single-shot measurements as opposed to continuous-valued voltage measurements. The second section builds in intensity with duplicated passages that are re-pitched and re-filtered to achieve a broader sonic palate.

The third section begins relatively calm, with more state-assigned measurement traces dictating changes in pitch of the *Bad Follower* synthesizer, which is tuned to have a bell-like tonality. The serenity is then disrupted by a droning note that slowly changes pitch (this is created by using measurements of offset charge as a function of time as a CV, as described in Section 3). This transition brings about an image of an approaching airplane disturbing the peace of a desolate landscape. The drone signal is then also used to modify the base pitch of the *2-state* synthesizer, where the continuous shift in pitch through microtones indicates clearly that something is changing, and not for the better. Suddenly, a short, staccato phrase (a perfect 4[th] interval, generated by a snippet from the *4-state* synth) sounds the alarm. Noise begins to reenter the sonic landscape, and the alarm sounds more frequently before the entrance of pulsed tones with increased harmonic content and ramped phase modulation, performed manually in a metaphorical battle between the instruments. At this point there is no semblance remaining of the calm from the start of the passage.

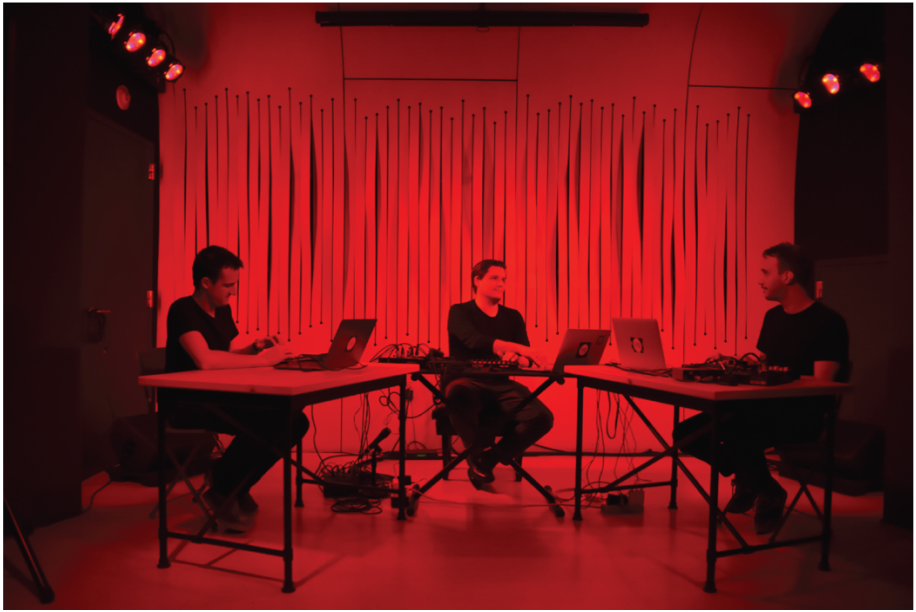

*Figure 9 - Photograph of the live performance at Firehouse 12 in the 3rd sonic section "Crescendo (melodic into danger)," with performers from left to right: Luke Burkhart, Spencer Topel, and Kyle Serniak.*

The opponents fade away as a serene atmosphere generated by the good-follower algorithm rises in amplitude. The plate reverb from Firehouse 12 gives the impression of retreat as one of the combatants lets out a final cry in the distance, signaling the beginning of the fourth and final passage.

Musical motifs from throughout the previous sections are reprised in this final section (excited-state "pings," manually-filtered noise, state-assigned pitch synthesis, and the slowly varying offset-charge drone, for example) with Luke's quantum control experiment laying a serene foundation. In the final moments, this tranquility is all that remains.

As a whole, the listener is encouraged to identify these sonic movements as metaphors for other aspects of experimental physics. For example, we had in mind a coarse representation of the cooldown procedure of a dilution refrigerator. It is at different points both calm and chaotic, eventually arriving to a state of dynamic equilibrium (at least until it's time to warm up and load in a new experiment). An alternative and equally valid comparison can be drawn to the experiences of a PhD student en route to graduation.

## 5  Conclusions

A primary goal of using transmon qubits as musical-signal generators was to "hear" what is happening inside these complex experimental systems. While quantum physics isn't exactly the most accessible topic, translating its concepts into audible music is a step toward reaching a broader audience. By creating this science outreach project with sound in mind, we sought a natural link between the listener and the quantum devices, and to offer a sonic interpretation of something that had not been heard before.

Shortly after the inception of the project, we realized that one potential challenge would be easily overcome: that of building a shared language between electronic music and experimental data. It boils down to the fact that signal processing is signal processing, regardless of whether you're working at audio frequencies or in the microwave range! By resampling the experimental signals to an audio rate, the signals were immediately interpretable in a musical context as control voltages and raw audio. This meant that not only were we able to quickly start playing with the raw data as sonic objects, but we were also able to consider new possibilities for control and sound synthesis. As we proceeded into the design of simple synthesizers, we drew inspiration from classic analog synthesizers which provided all the tools necessary for the translation of experimental data into audio signals.

While the synthesizers built here could be easily generated or simulated using classical systems, the fact that they are constructed from truly quantum systems opens a door to the utilization of uniquely quantum phenomena for audio synthesis. The player can control certain aspects of the musical creation (or more accurately, force the quantum system into a certain state to produce the type of sound desired), while the stochastic nature of the system, the possibility of errors, and the influence of the environment add extra dimensions inherent to the physical implementation of the synthesizer. The authorship of the music is shared with a truly quantum device. Extensions of this work could utilize quantum correlations in larger arrays of qubits as a step toward truly quantum musical synthesis. This work points toward future utility of prototype physical devices (quantum or otherwise) as tools for artistic and humanistic applications yet to be imagined.

# 6 Acknowledgments

The authors thank the Yale Quantum Institute for its financial support of this project, Michel Devoret and Robert Schoelkopf for their constant support and for graciously allowing us to use their labs' experimental systems, Firehouse 12 and their audio engineer Greg DiCrosta for hosting the live performance and the recording, the International Festival of Arts & Idea staff, Martha W. Lewis and the WPKN radio station for the broadcasting the live show, and Sang Wook Nam for mastering the album.